%% file: main.tex
\documentclass[runningheads]{llncs}

\usepackage[mobile]{eccv}

\usepackage{eccvabbrv}

\usepackage{graphicx}
\usepackage{epsfig}
\usepackage{amsmath}
\usepackage{amssymb}
\usepackage{caption}
\usepackage{booktabs, siunitx}
\usepackage{makecell} 
\usepackage[accsupp]{axessibility}  
\usepackage{mathtools}
\usepackage{url}

\newcommand{\defeq}{\coloneqq}

\newcommand{\E}{\mathbb{E}}

\newcommand{\Eb}[2]{\E_{#1}\!\left[#2\right]}

\DeclareMathOperator*{\argmin}{arg\,min}

\usepackage[pagebackref,breaklinks,colorlinks,citecolor=eccvblue]{hyperref}

\usepackage{orcidlink}
\usepackage{multicol}
\usepackage{multirow} 
\usepackage{wrapfig}
\usepackage{pifont}

\makeatletter
\def\blfootnote#1{\xdef\@thefnmark{}\@footnotetext{\scriptsize #1}}
\makeatother

\begin{document}

\title{InstructBrush: Learning Attention-based Instruction Optimization for Image Editing} 


\author{
Ruoyu Zhao \inst{1,2*} \and
Qingnan Fan \inst{2} \and
Fei Kou \inst{2} \and
Shuai Qin \inst{2} \and
Hong Gu \inst{2} \and
Wei Wu \inst{2,3} \and
Pengcheng Xu \inst{2,4} \and
Mingrui Zhu \inst{1} \and
Nannan Wang \inst{1\dagger} \and
Xinbo Gao \inst{5}
}

\titlerunning{InstructBrush}
\authorrunning{R. Zhao et al.}


\institute{
$^\text{1 }$Xidian University~~~$^\text{2 }$VIVO~~~
    $^\text{3 }$CityU~~~$^\text{4 }$Western University~~~ \\$^\text{5 }$Chongqing University of Posts and Telecommunications \\
\vspace{0.5em}
Project Page: \url{https://royzhao926.github.io/InstructBrush/}
}

\maketitle
\blfootnote{$*$Work done during the students' internships at VIVO. \\$\dagger$corresponding author.}

\begin{figure}[t]
  \centering
   \includegraphics[width=\textwidth]{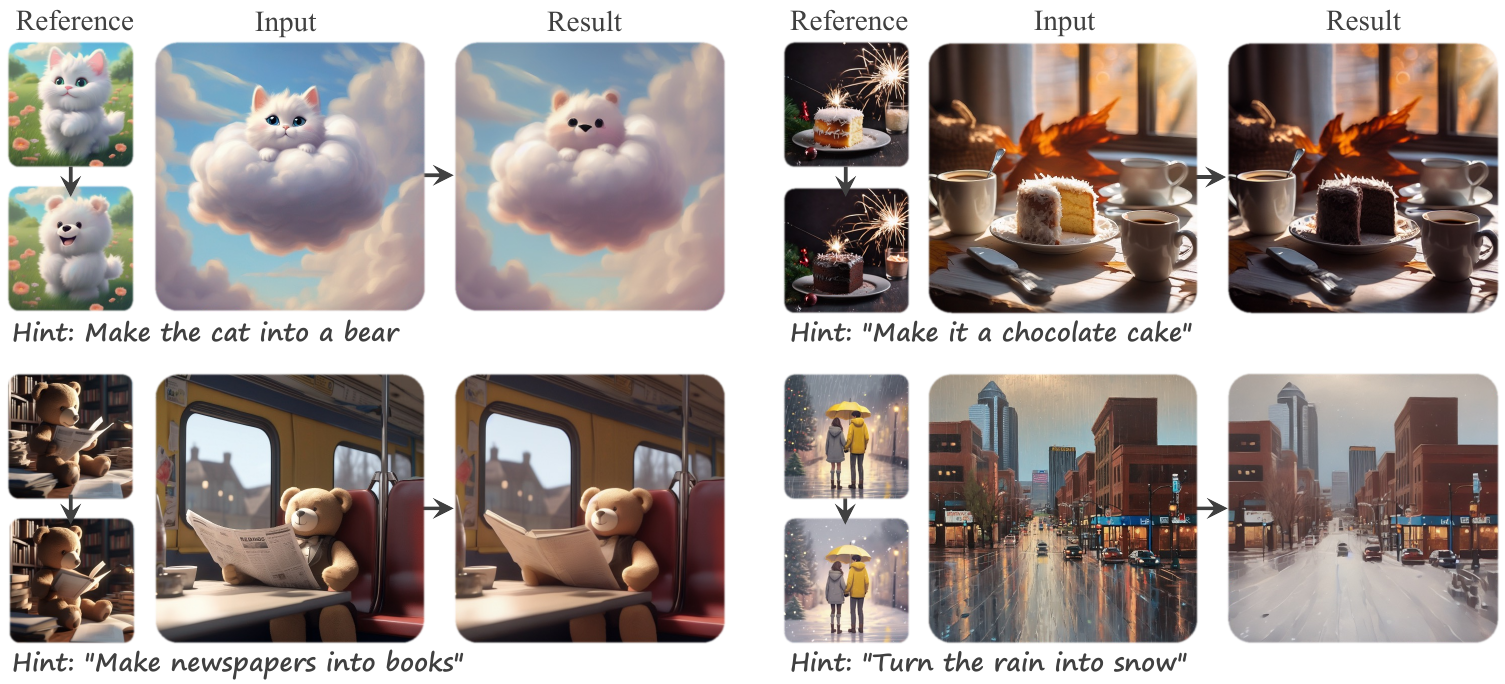}
   \caption{\textbf{\textit{InstructBrush}}: an inversion method for instruction-based image editing methods. It extracts editing effects from few reference image pairs as editing instructions, which are further applied for image editing. our InstructBrush achieves superior performance in editing and is more semantically consistent with the target editing effects.
   }
   \label{fig:teaser}
\end{figure}

\input{sections/abstract}
\input{sections/introduction}
\input{sections/related_work}
\input{sections/preliminaries}
\input{sections/method}
\input{sections/benchmark}
\input{sections/experiments}
\input{sections/conclusion}

\clearpage 

\bibliographystyle{splncs04}
\bibliography{egbib}

\appendix
\input{sections/appendix}

\end{document}

%% file: sections/abstract.tex
\vspace{-0.5em}
\begin{abstract}
In recent years, instruction-based image editing methods have garnered significant attention in image editing. However, despite encompassing a wide range of editing priors, these methods are helpless when handling editing tasks that are challenging to accurately describe through language. We propose \textbf{\textit{InstructBrush}}, an inversion method for instruction-based image editing methods to bridge this gap. It extracts editing effects from exemplar image pairs as editing instructions, which are further applied for image editing. Two key techniques are introduced into InstructBrush, \textit{Attention-based Instruction Optimization} and \textit{Transformation-oriented Instruction Initialization}, to address the limitations of the previous method in terms of inversion effects and instruction generalization. To explore the ability of instruction inversion methods to guide image editing in open scenarios, we establish a \textbf{T}ransformation-\textbf{O}riented \textbf{P}aired \textbf{Bench}mark (\textbf{TOP-Bench}), which contains a rich set of scenes and editing types. The creation of this benchmark paves the way for further exploration of instruction inversion. Quantitatively and qualitatively, our approach achieves superior performance in editing and is more semantically consistent with the target editing effects.
  \keywords{Image Editing \and Prompt Inversion \and Diffusion Models}
\end{abstract}

%% file: sections/introduction.tex
\section{Introduction}
\label{sec:intro}

Recently developed instruction-based image editing methods \cite{brooks2023instructpix2pix,zhang2023magicbrush,geng2023instructdiffusion,fu2023guiding,sheynin2023emu} enable users to effortlessly achieve their editing goals using natural language instructions. These methods have garnered significant attention owing to their flexibility and versatility in image editing, concurrently fostering research on a unified vision task framework \cite{geng2023instructdiffusion}. 
While instruction-based image editing models encompass rich prior knowledge, they still face challenges when dealing with editing that are difficult to express accurately through textual instructions. 
Therefore, exploring the replacement of textual instructions more consistent with the desired target transformation to guide the precise editing of new images is an important research problem.

This motivates the demand for the problem of \textit{instruction inversion}, which learns an instruction from a handful of image pairs exhibiting the same editing effect, and subsequently applies it to edit new images.  These image pairs that provide information about image transformations are also called \textit {visual prompts}. Instruction inversion not only serves as a valuable replacement when language is imprecise in describing specific editing concepts but also provides a promising solution for extracting priors in instruction-based editing models, aiding in the training of downstream vision tasks \cite{lee2023exploiting,zhao2023unleashing}.

The closest work to ours in this field is \cite{nguyen2023visual}, which attempts to invert visual prompts as text instructions for editing new images. It struggles with the editing effects for two reasons: 1) Inverting instructions in CLIP space limits their representational ability. Since CLIP is trained on text-image pairs with rough descriptions, it is challenging to provide specific representations of the image transformation details \cite{chen2023anydoor}. 2) Its instruction initialization strategy introduces editing-irrelevant content prior to instruction optimization, hence limiting the generalization of the instruction in generalized scenarios.

To bridge these gaps, we introduce \textit{\textbf{InstructBrush}}, a method to extract the editing effect from image pairs as editing instruction and then apply it to edit new images. In contrast to the previous method, we propose the \textit{Attention-based Instruction Optimization}. It localizes and optimizes the editing instruction within the cross-attention layers in the U-Net architecture \cite{ronneberger2015u} of the diffusion model, providing a more direct and effective approach to guide image editing. To assist in the optimization, we introduce the \textit{Transformation-oriented Instruction Initialization}. 
It ingeniously incorporates the editing-related prior into the learned instruction by identifying \textit{unique phrases} that describe the changes before and after image editing. This effectively mitigates the risk of compromising instruction generalization by introducing irrelevant content information and promotes semantic alignment of the instruction with the objectives.

To investigate the ability of the instruction inversion methods in guiding image editing in diverse scenarios, we establish \textbf{T}ransformation-\textbf{O}riented \textbf{P}aired \textbf{Bench}mark (\textbf{TOP-Bench}) tailored for instruction inversion tasks. This benchmark comprises a total of 750 images, encompassing 25 distinct editing effects, with each effect having 10 pairs of training data and 5 pairs of testing data. The creation of this benchmark not only helps to evaluate the potential of existing methods in guiding image editing, but also paves the way for further research in instruction inversion.
Qualitatively and quantitatively, our method surpasses the existing methods in terms of performance and demonstrates greater semantically consistency with the target editing effects.

In summary, our contributions are threefold:
\begin{itemize}
    \item We introduce \textit{\textbf{InstructBrush}}, a novel solution to instruction inversion, which extracts the editing instruction from exemplar image pairs for the subsequent image editing task.
    \item We propose the \textit{Attention-based Instruction Optimization} which is optimized within the feature space of the cross-attention, improving the results of instruction inversion, and the \textit{Transformation-oriented Instruction Initialization} for instructions optimization, ingeniously integrating textual priors capturing changes between images pairs into the optimized instructions.
    \item We establish \textbf{T}ransformation-\textbf{O}riented \textbf{P}aired \textbf{Bench}mark (\textbf{TOP-Bench}) for instruction inversion to assess its adaptability across diverse scenarios. Both qualitatively and quantitatively, our approach achieves more robust editing and is more semantically consistent with the target editing effects.
\end{itemize}

%% file: sections/related_work.tex
\section{Related Work}
\noindent{\textbf{Instruction-based Image Editing.}} 
Text-guided diffusion models \cite{nichol2021glide,ramesh2022hierarchical,saharia2022photorealistic,rombach2022high, podell2023sdxl,betker2023improving,dai2023emu} have taken the world by storm. By leveraging the robust generative priors of these models, 
InstructPix2Pix (IP2P) \cite{brooks2023instructpix2pix} makes the initial attempt to use a triplet dataset for training a model that edits images based on instructions, achieving intuitive and user-friendly instruction-based image editing. 
HIVE \cite{zhang2023hive} incorporates reward learning from human feedback to fine-tune IP2P for instruction editing that is more aligned with user preferences. To explore the application of instruction-based editing in real-world images, MagicBrush \cite{zhang2023magicbrush} constructs a large-scale manually annotated dataset to fine-tune IP2P, greatly improving the effect in real image editing.
Several existing methods, such as InstructDiffusion \cite{geng2023instructdiffusion} and Emu Edit \cite{sheynin2023emu} extend instruction-based editing methods to new visual tasks, demonstrating the potential of instruction-based image editing methods as a universal framework for visual tasks.
Recently, some efforts \cite{fu2023guiding,huang2023smartedit} leverage Multimodal Large Language Models (MLLMs) to enhance the performance of instructions, facilitating more accurate editing. Other efforts \cite{simsar2023lime,guo2023focus,li2023zone} concentration flexible and high-fidelity local editing, addressing the limitations of instruction-based editing in processing local details of images. Additionally, instruction-based image editing has been extended to 3D \cite{chen2023shap} and video \cite{xing2023vidiff} editing tasks, showcasing its tremendous application value.

\noindent{\textbf{Diffusion-based Prompt Inversion.}} 
The diffusion-based prompt inversion methods aim to learn the text prompt from a handful of images describing concepts, thereby guiding the generation of diffusion models. Textual Inversion \cite{gal2022image} learns text embeddings corresponding to pseudo-words to represent the target concepts. The pseudo-words can be combined with free text to guide the generation of images containing target concepts. Based on their research, some works  \cite{daras2022multiresolution,voynov2023p+,zhang2023prospect,alaluf2023neural,zhao2023catversion} explore the effects of different inversion spaces on prompt inversion. Other works \cite{gal2023encoder,wei2023elite,arar2023domain,chen2023photoverse,ye2023ip,li2023photomaker} train an image encoder based on text inversion to achieve generation guided by a given reference image. Additionally, 
ReVersion \cite{huang2023reversion} focuses on learning the relation between objects through contrastive learning. 
PEZ \cite{wen2023hard} inverts hard prompts by projecting learned embeddings onto adjacent interpretable word embeddings, providing a new solution for image captioning.  
\cite{vinker2023concept} decomposes a visual concept, allowing users to explore hidden sub-concepts of the object of interest. 
Lego \cite{motamed2023lego} uses carefully designed prompt learning methods to learn abstract concepts that are entangled with the subject from few samples. These methods focus on learning concepts to guide image generation, while our study aims to learn the transformations between image pairs to guide image editing.

\noindent{\textbf{Visual In-context Learning.}} 
In-context learning \cite{brown2020language}, which originated from the field of natural language processing (NLP), has been promoted as a learning paradigm. This paradigm enables the execution of a given task on a sample query after learning the task from a set of examples. VisualPrompting \cite{bar2022visual} first introduced the concept of visual contextual learning. It uses an inpainting-based approach with grid-like inputs and has shown remarkable results in many tasks. Subsequent works \cite{wang2023images,wang2023seggpt,fang2024explore} broaden the application areas of the framework, such as keypoint detection \cite{wang2023images}, image denoising \cite{wang2023images}, image segmentation \cite{wang2023seggpt} and 3D point cloud \cite{fang2024explore}. 
Recent works \cite{wang2024context,chen2023improving} introduce in-context learning on diffusion models to accomplish various visual tasks, but they require guidance from textual instructions. ImageBrush \cite{yang2024imagebrush} models visual transformations as a diffusion-based inpainting problem. However, it still requires grid images as input, which poses a significant burden when processing high-resolution images. Unlike these methods, Visii \cite{nguyen2023visual} focuses on editing tasks. It inverses exemplar image pairs into a text instruction within an instruction-based image editing model, replacing textual instructions to guide the editing of new images. Our approach similarly focuses on image editing based on instruction inversion and achieves more robust editing and generalization ability to new scenarios. 

\begin{figure}[t]
  \centering
\includegraphics[width=1\textwidth]{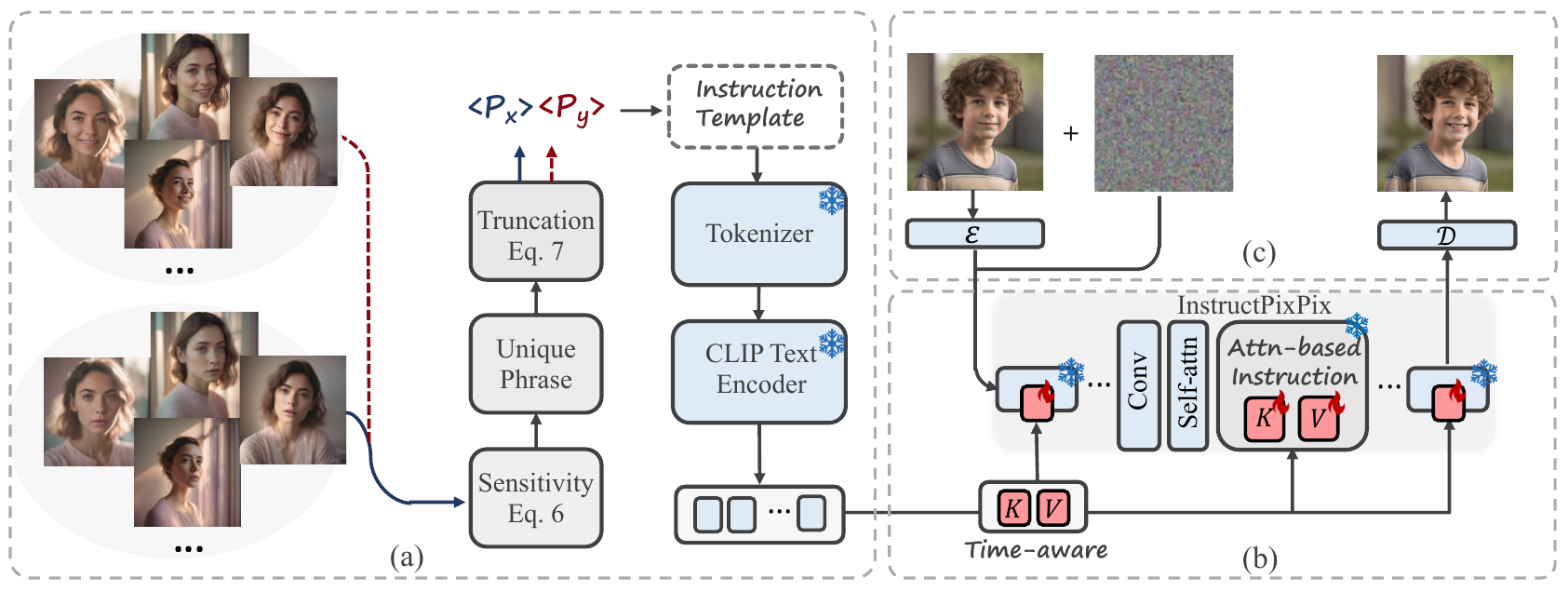}
   \caption{\textbf{The Framework of InstructBrush}. InstructBrush inverts instructions from exemplar image pairs by proposing novel \textit{Transformation-oriented Instruction Initialization} (a) and \textit{Attention-based Instruction Optimization} (b) modules. 
   The former is proposed to initialize the instruction, which effectively introduces the editing-related prior to facilitate semantic alignment of the instruction with the exemplar image pairs. 
   The latter introduces the editing instruction into the cross-attention layers of the instruction-based image editing model and directly optimizes the Keys and Values corresponding to the instruction within these layers.
   After optimization, the learned instructions are used to guide the editing of new images (c).
   }
   \label{fig:framework}
\end{figure}

%% file: sections/preliminaries.tex
\section{Preliminaries}
\noindent{\textbf{Latent Diffusion Models.}} 
Stable Diffusion (SD), a variant of the latent diffusion model (LDM) \cite{rombach2022high}, serves as a text-guided diffusion model. To generate high-resolution images while enhancing computational efficiency in the training process, it employs a pre-trained variational autoencoder (VAE) encoder $\mathcal{E}(\cdot)$ to map images into latent space and perform an iterative denoising process. Subsequently, the predicted images is mapping back into pixel space through the pre-trained VAE decoder $\mathcal{D}(\cdot)$. For each denoising step, the simplified optimization objective is defined as follows:
\begin{gather}
    L_{LDM}(\theta) \defeq \Eb{\mathcal{E}(x), \epsilon, t} { { \left\| \epsilon - \epsilon_\theta(z_t, t, \tau_\theta(c) \right\|^2_2}}. \label{eq:ldm_loss}
\end{gather}
In this process, the text description $c$ is first tokenized into textual embeddings by a Tokenizer. The textual embeddings are then passed through the CLIP text encoder $\tau_\theta(\cdot)$ to obtain text conditions. The resulting text conditions are used to guide the diffusion denoising process.

\noindent{\textbf{InstructPix2Pix.}} Our method is based on InstructPix2Pix (IP2P) \cite{brooks2023instructpix2pix}, an instruction-guided image editing method. After encoding the input image $c_I$ using the VAE encoder, IP2P concatenates the noisy latent $z_t$ with the encoded latent $\mathcal{E}(c_I)$ in the first convolutional layer of SD. Subsequently, it uses a generated triplet dataset to perform instruction tuning \cite{wei2021finetuned} on the improved network. This method maximizes the utilization of SD's powerful generative prior, thereby enabling stunning image editing based on human instructions $c_T$. The simplified denoising optimization objective is defined by:
\begin{gather}
    L_{IP2P}(\theta) \defeq \Eb{\mathcal{E}(x), \mathcal{E}\left(c_I\right), c_T, \epsilon, t}{ { \left\| \epsilon - \epsilon_\theta(z_t, t, \mathcal{E}(c_I), c_T) \right\|^2_2}}. \label{eq:ip2p_loss}
\end{gather}
The dual conditional framework of IP2P employs both input image $I$ and text instruction $t$ for guidance, achieved through an enhanced classifier-free guidance (CFG) strategy \cite{ho2022classifier}. The improved CFG incorporates two distinct guidance scales, $s_T$ and $s_I$, adjustable to balance guidance strength between text and image conditions. It learns the score estimate predicted by the network corresponding to a single denoising step as follows:
\begin{equation}
\begin{aligned}
\tilde{e}_\theta\left(z_t, c_I, c_T\right) &= e_\theta\left(z_t, \varnothing, \varnothing\right) \\
&\hphantom{=} + s_I \cdot \left(e_\theta\left(z_t, c_I, \varnothing\right) - e_\theta\left(z_t, \varnothing, \varnothing\right)\right) \\
&\hphantom{=} + s_T \cdot \left(e_\theta\left(z_t, c_I, c_T\right) - e_\theta\left(z_t, c_I, \varnothing\right)\right). \label{eq:ip2p_cfg}
\end{aligned}
\end{equation}

\begin{figure}[t]
  \centering
\includegraphics[width=1\textwidth]{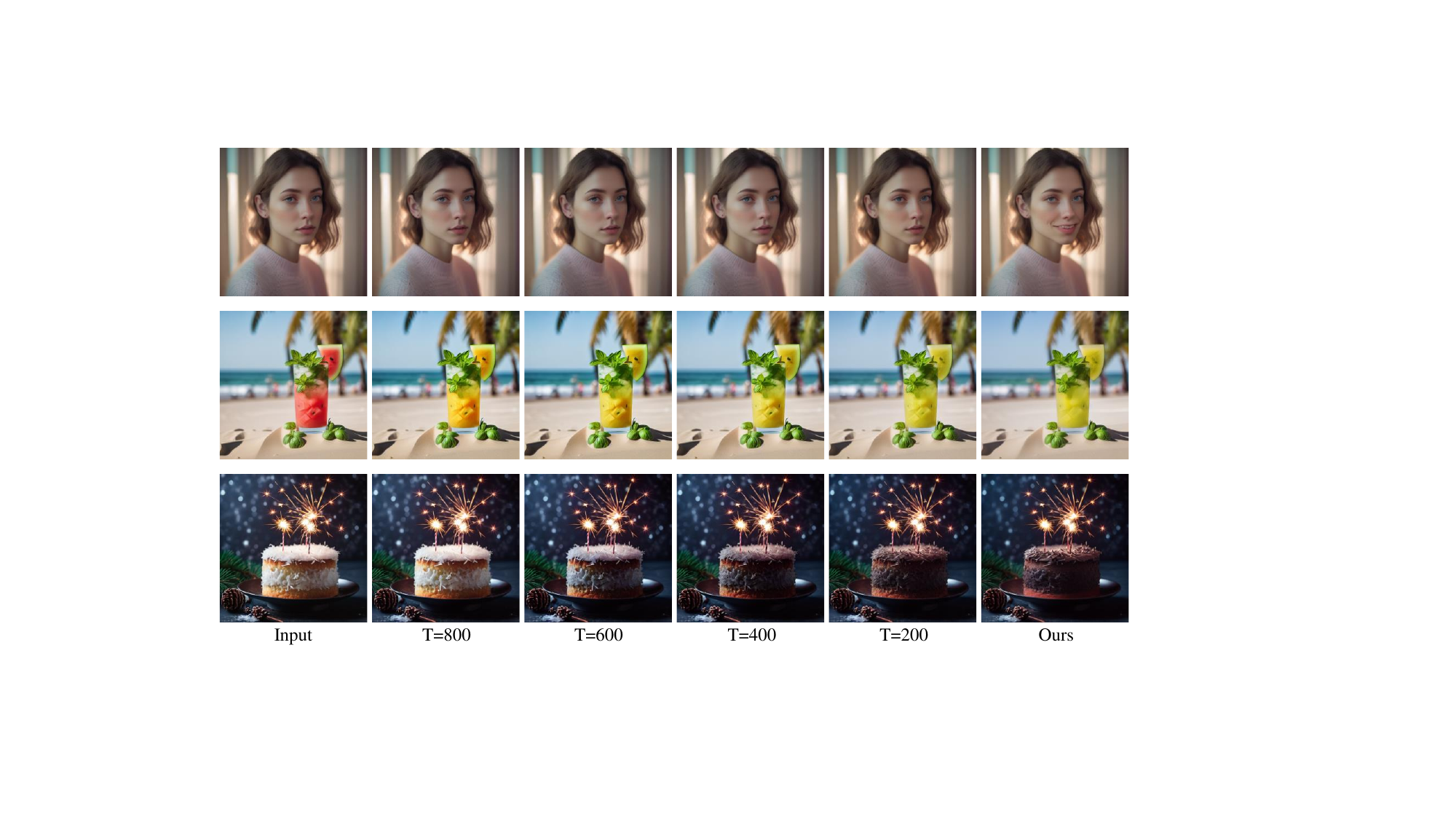}
   \caption{\textbf{Visualization of Applying Time-aware Instructions to Various Denoising Steps}. Example: $T = 800$ represents the application of our time-aware instruction before the denoising time step of 800 (steps 1000 to 800), while the \textit{None} instruction is applied to the denoising process after 800 steps (steps 800 to 0). Therefore, $T = 1000$ indicates the input image, and $T = 0$ indicates our full implementation. The visualization results show that in the early denoising stages, the editing focuses on coarse information such as colors (rows 2 and 3); in the later stages, the editing focuses on detailed information such as textures and facial expressions (rows 1 and 3).
   }
   \label{fig:timeaware}
\end{figure}

%% file: sections/method.tex
\section{Method}
\label{sec: Method}
The pipeline of \textit{InstructBrush} is demonstrated in Figure \ref{fig:framework}. Based on the instruction-based image editing methods \cite{brooks2023instructpix2pix}, \textit{InstructBrush} inverts exemplar image pairs as editing instructions and applies them to editing new images. It proposes novel \textit{Attention-based Instruction Optimization} and \textit{Transformation-oriented Instruction Initialization} modules. The former introduces the editing instruction into the cross-attention layers of the instruction-based image editing model and directly optimizes the Keys and Values corresponding to the instruction within these layers, facilitating more effective instruction inversion (Section~\ref{sec: instruction optimization}). The latter is proposed to initialize the optimized instruction, which effectively introduces the editing-related prior to facilitate semantic alignment of the instruction with the exemplar image pairs (Section~\ref{sec: instruction initialization}).

\subsection{Attention-based Instruction Optimization}
\label{sec: instruction optimization}
Inspired by Textual Inversion \cite{gal2022image}, The current instruction inversion method \cite{nguyen2023visual} optimizes the output embeddings of the CLIP text encoder using image pairs, aiming to represent the transformation effects between image pairs in CLIP space. 
However, CLIP is trained on text-image pairs with rough descriptions, and its feature space is prone to losing the detailed representation of the image \cite{chen2023anydoor}. Therefore, it is difficult to achieve the requirement of only optimizing the instruction that represents the target transformation in CLIP space.
Instead, we focus on optimizing the features in cross-attention. These features are projected from textual embeddings to representations consistent with image features, enabling a more precise representation of image transformation details \cite{hertz2022prompt,simsar2023lime}.
As a result, we introduce an attention-based instruction optimization that optimizes editing instructions in the image feature space of the cross-attention layers in the diffusion model, fostering more effective instruction inversion.

\noindent{\textbf{Attention-based Instruction.}}
Considering a single-head cross-attention, let $Q$ be the query, $K$, $V$ be the keys and values from the instruction, respectively, the cross-attention is given by:
\begin{gather}
\begin{aligned}
    \text{Attention}\left(Q, K, V\right)= \text{Softmax}\Big(\frac{QK^T}{\sqrt{d'}} \Big)V.
\end{aligned}\label{eq:cross_attention}
\end{gather}
Here, $K, V \in \mathbb{R}^{l \times d}$, where $l$ represents the token length of the instruction, and $d$ represents the feature dimension, the value of which depends on the position of the cross-attention layer in the U-Net framework. We optimize the features $\gamma_K, \gamma_V \in \mathbb{R}^{m \times d}$ with a length of $m \in l$ in the key and value corresponding to the first $m$ tokens of the text instruction. Because after linear projection, instruction embeddings transform from text embedding to image features, exhibiting stronger image representation capabilities. 
To optimize the feature embeddings of the editing instruction, our optimization objective is derived from the simplified least squares error in Eq.~\ref{eq:ip2p_loss}:

\begin{gather}
\begin{aligned}
    \gamma = \argmin \Eb{\mathcal{E}(x), \mathcal{E}\left(c_I\right), c_T, \epsilon, t}{ { \left\| \epsilon - \epsilon_\theta(z_t, t, \mathcal{E}(c_I), c_T) \right\|^2_2}}.
\end{aligned}\label{eq:total_loss}
\end{gather}
Here, $\gamma = \left\{ \gamma_K, \gamma_V \right\}_{1 \ldots n}$ represents the features of keys and values from the first $m$ tokens of the text instruction in all $n$ cross-attention layers. The value of $m$ corresponds to the number of text tokens used for instruction initialization, as described in Section~\ref{sec: instruction initialization}.

\noindent{\textbf{Time-aware Instruction.}}
In the text-guided diffusion models, the denoising process focuses on image generation from low-frequency structure to high-frequency details \cite{daras2022multiresolution,zhang2023prospect}. 
We believe that a similar property also exists in instruction-based editing models, where different denoising processes primarily focus on distinct transformations. Therefore, we divide the instruction optimization equally into $j$ parts based on denoising time steps, allowing instructions to focus on different editing tasks at different denoising time steps. We visualize the effect of applying our time-aware instructions on image editing during different denoising periods in Figure \ref{fig:timeaware}. The visualization results show that in the early denoising stages, the editing focuses on coarse information such as colors; in the later stages, the editing focuses on detailed information such as textures and facial expressions.
Now, we have $\gamma = \left\{ \gamma_K, \gamma_V \right\}^{j}_{1 \ldots n}$, where $j$ is 5 by default.
In this way, the learned instructions can capture more details of transformations, which can guide the editing of new images more robustly.

\subsection{Transformation-oriented Instruction Initialization} 
\label{sec: instruction initialization}
Concept inversion (such as Textual Inversion), using the semantic class word (\textit{e.g.}, dog, cat) for initialization, provides prior information for the target concept learning. However, the instruction inversion requires learning a sentence describing the \textit{image transformation}. Manually initializing a sentence based on the transformation of reference image pairs is not only laborious but also subjective. 
The existing method \cite{nguyen2023visual} utilizes the caption method \cite{wen2023hard} to obtain the caption of after-editing images in the training set as the start point of the optimization. Despite the introduction of transformation-related prior knowledge, this content-oriented initialization approach can simultaneously introduce editing-irrelevant content information about the training scenario, hindering the generalization of instruction to new scenarios.
Instead, our method directly extracts the information related to image transformations that assist in the optimization process of instruction without affecting its generalization ability. Specifically, we first extract \textit{unique phrases} that differentiate the images before and after editing as editing-related priors. Subsequently, we incorporate them into the \textit{image transformation template} for instruction initialization.

\noindent{\textbf{Unique Phrase Extraction.}}
Given a set of image pairs $\{\{x\}, \{y\}\}$, where $\{x\}$ and $\{y\}$ represent the image sets before and after editing, for a single set $\{x\}$, we search for a caption based on the image-text feature similarity as $CAP_x=\{<\!p_1\!>, \ldots, <\!p_r\!>\}$. Here, $<\!p_i\!>$ represents a phrase from the vocabulary of \cite{pharmapsychotic-clip-interrogator}, and $r$ represents the adjustable number of phrases to form the caption, which is set to 5 by default. 
Subsequently, we compare the feature similarity between $CAP_x$ and the image sets $\{x\}$, $\{y\}$ respectively, and then measure the difference in feature similarity of the same phrase with the two sets as the \textit{sensitivity}. This process can be represented as follows:
\begin{gather}
\begin{aligned}
    sens_i\left(<\!p_i\!> \right)=sim\left(<\!p_i\!>, \{x\}\right)-sim\left(<\!p_i\!>, \{y\}\right)
\end{aligned}\label{eq:unique_phrase}
\end{gather}
Here, $sens_i$ denotes the sensitivity of the $i$th phrase in $CAP_x$ and $sim$ denotes the CLIP feature similarity.
We identify the phrase with maximum sensitivity as the \textit{unique phrase} $<\!p_x\!>$ of the set $\{x\}$.
However, there exist certain edits whose relevant information cannot be recognized by a caption. To avoid the unique phrase containing editing-irrelevant information, we define the following conditions:
\begin{gather}
\begin{aligned}
    <\!p_x\!> = \begin{cases}<\!p_x\!> & \text { if }sens\left(<\!p_x\!>\right) \geq \eta \\ \varnothing & \text { otherwise},\end{cases}
\end{aligned}\label{eq:phrase_condition}
\end{gather}
where $\eta$ represents a constant that controls the truncation of unique phrases, set to 0.15 by default.

\noindent{\textbf{Image Transformation Template.}}
With the above method, we can get the unique phrases $<\!p_x\!>$ and $<\!p_y\!>$ for sets $\{x\}$ and $\{y\}$. Then we incorporate them into the image transformation template. For example, we use $''turn <\!p_x\!> into <\!p_y\!>''$ as a starting point for instruction optimization. Note that when $<\!p_y\!> = \varnothing$, we use \textit{None} instruction for initialization and optimize fixed-length features for Keys and Values in cross-attention. Although the initialized instruction is not sufficient to express the target editing effect, it can introduce prior knowledge of transformation, aiding the semantics of learned instruction to be close to the target.

%% file: sections/benchmark.tex
\section{Transformation-oriented Paired Benchmark}
\label{sec:benchmark}
To investigate the editing capabilities of various instruction inversion methods in open scenarios and facilitate a fair comparison of these methods, We establish a benchmark named \textit{TOP-Bench} (\textbf{T}ransformation-\textbf{O}riented \textbf{P}aired \textbf{Bench}mark), which can be utilized for both qualitative and quantitative evaluations. 
Our benchmark spans 25 datasets corresponding to different editing effects. It covers a wide range of editing categories and scenarios, allowing for division from multiple dimensions. Each dataset consists of 10 pairs of training images and 5 pairs of testing images, totaling 750 images. Additionally, we provide text instructions aligned with the transformation effects for each dataset. Please refer to the Supplementary for data acquisition and detailed introduction.

To further analyze the advantages of our method, we categorize the benchmark into two different categories: TOP-Global and TOP-Local, corresponding to datasets of 14 global editing effects and 11 local editing effects, respectively. We compare the quantitative results of different methods in these two categories to validate the effectiveness of our method.

%% file: sections/experiments.tex
\section{Experiments}
In this section, we present qualitative and quantitative results. The experimental results demonstrate that our proposed \textit{InstructBrush} outperforms existing methods.

\noindent \textbf{Implementation Details}.
The implementation is based on one NVIDIA Tesla V100 GPU. We use pre-trained IP2P and optimized the features of the keys and values corresponding to the instruction tokens (around 10) obtained from our instruction initialization method. We divide the learned instructions into 5 parts according to the denoising time step, and optimize each part with 1000 steps using a learning rate of 0.001 and a batch size of 1, respectively, for a total of 5000 steps. The whole training process takes about 25 minutes. 
During both training and inference, we adopt a text guidance scale $s_T=7.5$ and an image guidance scale $s_I=1.5$. And we use the Euler ancestral sampler with denoising variance schedule \cite{karras2022elucidating} with a sampling step of $T=20$ during the inference process.

\noindent \textbf{Metrics}.
We use four objective evaluation metrics on the benchmark. Specifically, we employ full-reference quality metrics PSNR, SSIM \cite{wang2004image}, and LPIPS \cite{zhang2018unreasonable} to assess the consistency between the generated images and the ground truth, quantifying the image editing capabilities of each method. In addition, we measure the CLIP directional similarity \cite{gal2022stylegan} between image pairs to evaluate the semantic alignment between the editing direction of each method and the target. Specifically, we measure the consistency between the average editing direction from the input images to the generated images and the average direction of the training image pairs.

\noindent \textbf{Compared Methods}.
We compare our \textit{InstructBrush} with the state-of-the-art competitor Visii \cite{nguyen2023visual} and the base editing model IP2P \cite{brooks2023instructpix2pix}. 
We use an image resolution of $512\times512$ for comparison with other methods. For Visii, we utilize its official implementation, while for IP2P, we employ its Diffusers \cite{von-platen-etal-2022-diffusers} version. All experiments are conducted following the official recommended configurations.

\begin{figure}[!ht]
  \centering\includegraphics[width=1\textwidth]{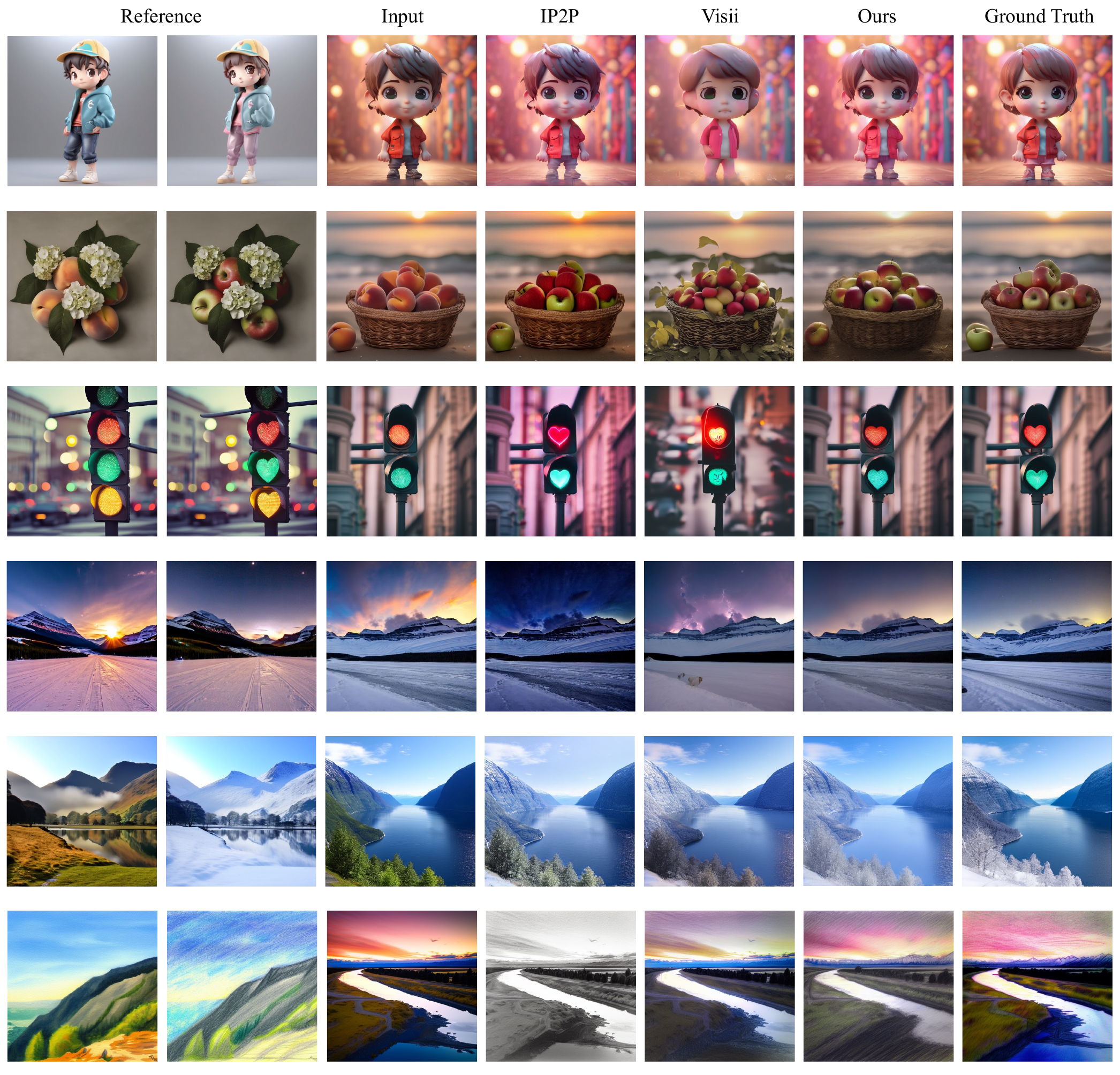}
   \caption{\textbf{Qualitative Comparisons with Existing Methods.} Our method achieves superior performance in both local and global image editing. It effectively avoids introducing editing-irrelevant information from the training images, showing better instruction generalization.
   }
   \vspace{-1em}
   \label{fig:method_comparison}
\end{figure}

\begin{figure}[!htb]
  \centering\includegraphics[width=1\textwidth]{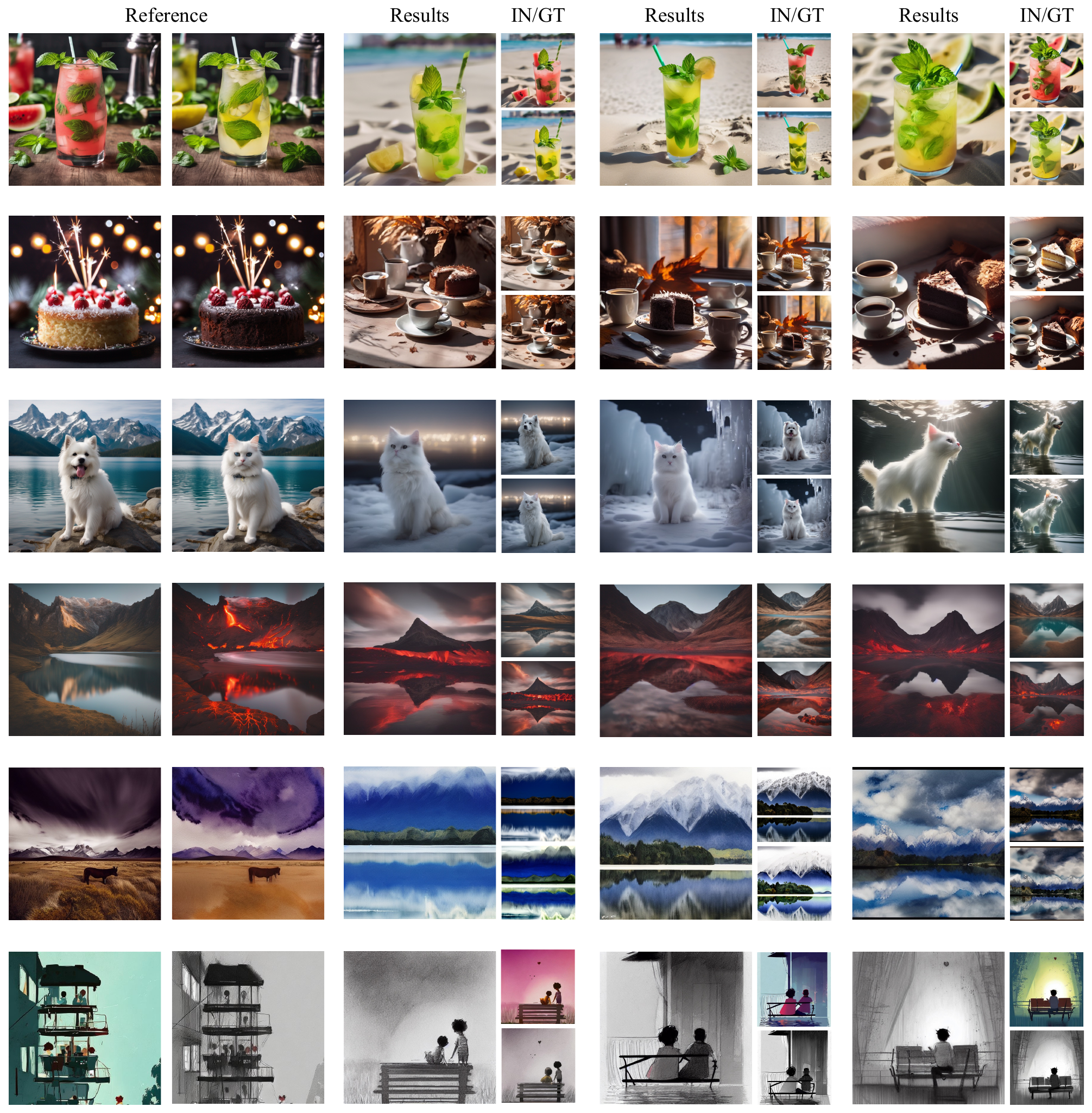}
   \caption{\textbf{More Visualization Results of Our Method.} Our method demonstrates robust performance on both local and global editing. And it does not introduce scene information of the training image when editing new images, which reflects the instruction generalization of our method.
   }
   \label{fig:our_results}
\end{figure}

\begin{table}[t]
\centering
\caption{\textbf{Quantitative Results.} We measure the average PSNR, SSIM, LPIPS, and CLIP direction scores of several methods in different editing tasks. Our method is significantly superior to other methods. We highlight in red the percentage of our method that exceeds Visii.}
\resizebox{0.8\columnwidth}{!}{
  \begin{tabular}{clllll}

    \toprule
    Datasets \quad  & Method \quad\quad & PSNR $\uparrow$ & SSIM $\uparrow$  & LPIPS $\downarrow$ & Direct. $\downarrow$ \\ 
    \midrule
    \multirow{3}*{TOP-Global\quad} & IP2P \cite{brooks2023instructpix2pix} \quad\quad  & 16.10 & 0.5467 & 0.2810 & 0.2997 \\ 
                      & Visii \cite{nguyen2023visual} \quad\quad & 15.87 & 0.4947 & 0.3866 & 0.3938 \\ 
                      & \textbf{Ours} \quad\quad & \textbf{18.66}\textcolor{red}{$_{17.5\%\uparrow}$} & \textbf{0.5842}\textcolor{red}{$_{18.1\%\uparrow}$} & \textbf{0.2526}\textcolor{red}{$_{34.7\%\downarrow}$} & \textbf{0.2798}\textcolor{red}{$_{28.9\%\downarrow}$}\\ 
    \midrule 
    \midrule
    \multirow{3}*{TOP-Local\quad} & IP2P \cite{brooks2023instructpix2pix} \quad\quad  & 20.12 & 0.7701 & 0.1648 & 0.5236 \\ 
                      & Visii \cite{nguyen2023visual} \quad\quad & 18.76 & 0.7157 & 0.2585 & \textbf{0.2560} \\ 
                      & \textbf{Ours} \quad\quad & \textbf{23.26}\textcolor{red}{$_{24.0\%\uparrow}$} & \textbf{0.8297}\textcolor{red}{$_{15.9\%\uparrow}$} & \textbf{0.1143}\textcolor{red}{$_{55.8\%\downarrow}$} & 0.3576\\ 
    \midrule  
    \midrule
    \multirow{3}*{TOP-Bench\quad} & IP2P \cite{brooks2023instructpix2pix} \quad\quad  & 17.87 & 0.6450 & 0.2298 & 0.3982 \\ 
                      & Visii \cite{nguyen2023visual} \quad\quad & 17.14 & 0.5919 & 0.3303 & 0.3332 \\ 
                      & \textbf{Ours} \quad\quad & \textbf{20.68}\textcolor{red}{$_{20.6\%\uparrow}$} & \textbf{0.6922}\textcolor{red}{$_{16.9\%\uparrow}$} & \textbf{0.1918}\textcolor{red}{$_{41.9\%\downarrow}$} & \textbf{0.3140}\textcolor{red}{$_{5.7\%\downarrow}$}\\ 
    \bottomrule
  \end{tabular}
}
\label{tab:compare_results}
\end{table}

\subsection{Comparisons}
\noindent \textbf{Qualitative Comparisons.}
To demonstrate the advantages of our \textit{InstruchBrush} in image editing, we compared it with the existing instruction inversion method Visii and the base model IP2P. We use our \textit{TOP-Bench} to evaluate the results of different methods. For instruction inversion methods, we employ 10 reference before-and-after editing image pairs to optimize the instruction for each editing effect. For IP2P, we utilize natural text instructions provided by \textit{TOP-Bench} for editing. Subsequently, we present a comparison of the results of editing the test images in Figure \ref{fig:method_comparison}. IP2P can easily edit input images using text instructions, but it is difficult to reproduce the editing effects displayed by images that are difficult to express in language. Even if we have text descriptions paired with reference images, due to slight deviations between the text instructions of the base model and the editing task \cite{sheynin2023emu}, there are failures and coupling situations, as shown in rows 3 through 6. Although Visii optimizes instructions to learn the target editing concept and solves the problem of IP2P not being able to specifically represent image changes using text instructions alone, its content-oriented initialization reduces the instructions generalization. It can easily introduce content information in the training image during the instruction editing process, as shown in rows 2 and 3. In addition, the limitations of optimization space also make it difficult to accurately learn target editing concepts. 
By contrast, our \textit{InstructBrush} demonstrates superior editing performance. Fig.~\ref{fig:our_results} illustrates more qualitative results obtained by our method. It demonstrates robust performance on both local and global editing. And it does not introduce scene information of the training image when editing new images, which reflects the instruction generalization of our method.

\noindent \textbf{Quantitative Comparisons.}
In addition to qualitative comparisons, we conduct a detailed quantitative evaluation of these methods on TOP-Local, TOP-Global, and overall TOP-Bench. As shown in Table \ref{tab:compare_results}, the editing performance of our method surpasses that of other methods at both the editing effects and semantic alignment. In addition, compared to the results of Visii, our method shows a more significant improvement on TOP-Local than on TOP-Global. This is because in local editing tasks, training images contain more editing-irrelevant scene information. The content-oriented initialization of Visii introduces them to the initialized instructions, posing a greater obstacle to optimization. On the contrary, the transformation-oriented instruction initialization used by our method can accurately capture the transformations between image pairs and use them for initialization, thus improving instruction generalization.

\begin{figure}[t]
  \centering\includegraphics[width=\textwidth]{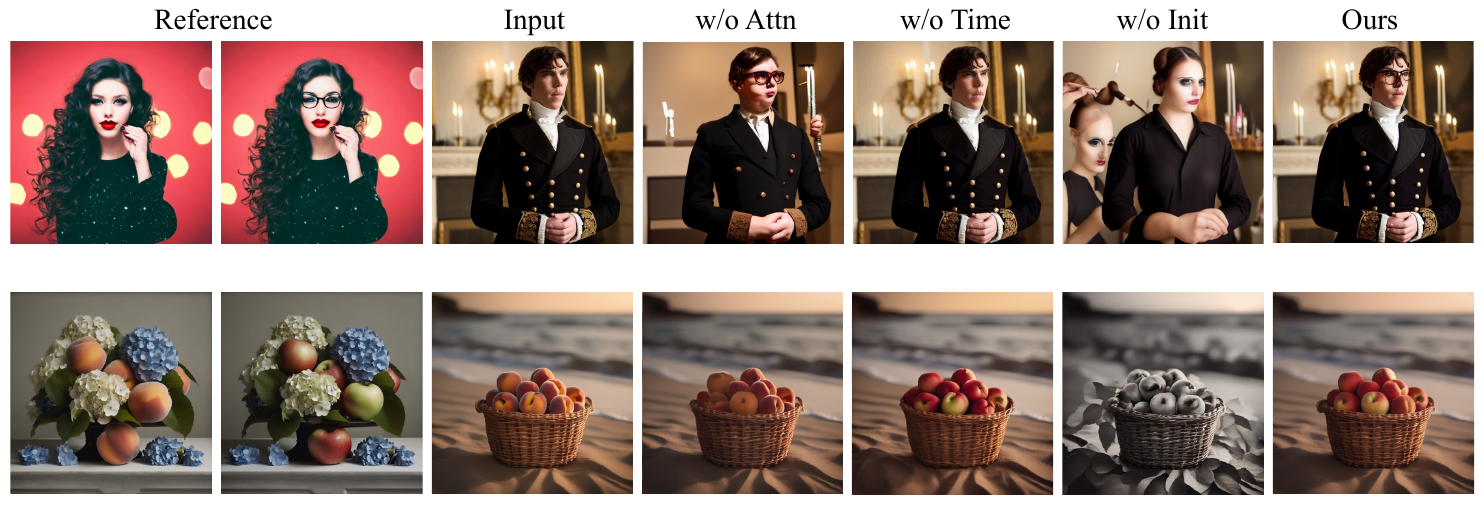}
   \caption{\textbf{Visualization Results of Ablation Study}. We visualize the independent effects of our proposed attention-based instruction, time-aware instruction, and transition-oriented instruction initialization on the results, intuitively highlighting the importance of these configurations.
   }
   \label{fig:ablation}
\end{figure}

\begin{table}[t]
\centering
\caption{\textbf{Ablation Study}. We validate the independent impact of our proposed attention-based instruction, time-aware instruction, and transformation-oriented instruction initialization on results, emphasizing the importance of these configurations.}
\resizebox{0.5\columnwidth}{!}{
  \begin{tabular}{cllll}
    \toprule
    Method & PSNR $\uparrow$  & SSIM $\uparrow$  & LPIPS $\downarrow$ & Direct. $\downarrow$ \\ 
    \midrule
    w/o Attn  & 19.56 & 0.6709 &0.2179 & 0.3124 \\ 
    w/o Time &19.64 & 0.6656 & 0.2271 & \textbf{0.2910} \\ 
    w/o Init & 20.22 & 0.6841 & 0.2018 & 0.3747 \\ 
    \textbf{Ours} & \textbf{20.68} & \textbf{0.6922} & \textbf{0.1918} & 0.3140\\ 
    \bottomrule
  \end{tabular}
}
\label{tab:ablation}
\end{table}

\subsection{Ablation Studies}

\noindent \textbf{Attention-based Instruction Ablation.}
The use of attention-based instruction aims to avoid the limitation of CLIP space on the representation ability of target transformations and achieve a more accurate representation of image transformation details. The metrics PSNR, SSIM, and LPIPS are calculated between the output and the ground truth to evaluate the editing performance. We report results in Table \ref{tab:ablation} and observe that adopting attention-based instructions replaced with CLIP space-based instructions effectively improves the editing performance of the instructions. Additionally, we also observe in Figure \ref{fig:ablation} that compared to inversion in CLIP space, optimizing instruction in attention space has shown significant improvements in editing.

\noindent \textbf{Time-aware Instruction Ablation.}
The use of time-aware instructions facilitates instruction optimization by allowing instructions to focus on learning different transformations at different denoising time steps. Table \ref{tab:ablation} explicitly shows that the use of time-aware instruction helps to improve the editing effect. The same result is confirmed in Figure \ref{fig:ablation}. This suggests that it is necessary to optimize the corresponding instructions for different denoising time steps to fit the image transform of interest for the current step.

\noindent \textbf{Transformation-oriented Instruction Initialization Ablation.}
Content-oriented initialization methods introduce irrelevant content information from the training images, thereby interfering with the optimization process. As depicted in Figure \ref{fig:ablation}, the use of the content-oriented initialization method results in the leakage of content information from the training image into the edited image. By enabling instruction initialization to prioritize image changes over image content, it not only enhances the editing capabilities of learned instructions, but also aligns the edited image with the target transformation in terms of semantic information, which is confirmed in Table \ref{tab:ablation}.

\setlength{\intextsep}{-5pt}
\begin{wrapfigure}{r}{0.3\textwidth}
  \includegraphics[width=0.3\textwidth]{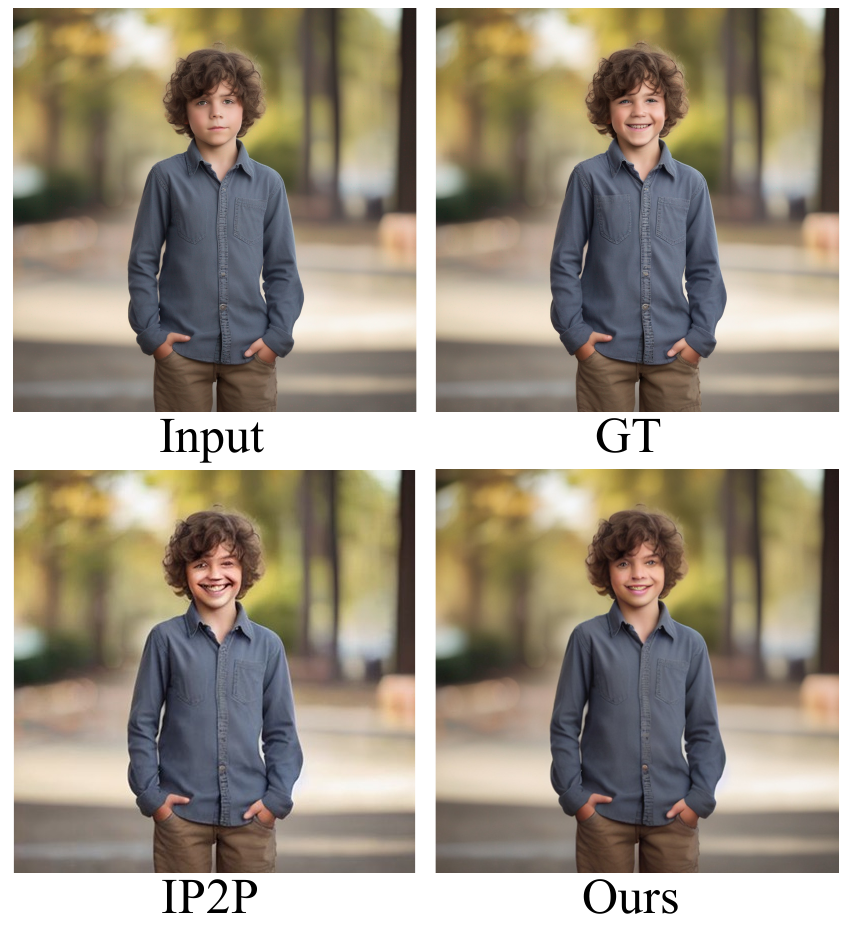}
  \caption{Failure case for our method.}
  \label{fig:fail}
\end{wrapfigure}
\subsection{Limitations}
As the implementation of this method relies on instruction-based editing models, the implementation of editing is constrained by the prior of the base model. As shown in Figure \ref{fig:fail}, even if we capture the editing concept, however, the upper bound for modifying subtle facial expressions still depends on the generation ability of IP2P.
In addition, our initialization method is limited by the vocabulary used to search for unique phrases. If the phrase is not present in the vocabulary, our initialization method will initialize using \textit{None} instruction, which will not introduce any editing prior.

%% file: sections/conclusion.tex
\section{Conclusion}
We present a novel approach to extracting the transformation effects accurately of image pairs into editing instructions, which are then utilized to guide the editing of new images. To enhance the accuracy and generalization of inverted instructions, we propose attention-based instruction optimization and transformation-oriented instruction initialization. In addition, we establish a benchmark for instruction inversion to further investigate the capabilities of various instruction inversion methods. We qualitatively and quantitatively validate the effectiveness of our proposed method. In the future, we will apply our method to more powerful instruction-based image editing models for more robust editing performance. We hope that this work will stimulate more research on instruction inversion and serve as a prior extraction method to help downstream task training.

%% file: sections/appendix.tex
\newpage
\setcounter{page}{1}

\section{Evaluaion Metrics}
\label{supp:evaluation_metrics}
We use four objective evaluation metrics on the benchmark. Specifically, we employ full-reference quality metrics PSNR, SSIM \cite{wang2004image}, and LPIPS \cite{zhang2018unreasonable} to assess the consistency between the generated images and the ground truth, quantifying the image editing capabilities of each method. Among them, higher PSNR indicates more similarity between the results and the ground truth; higher SSIM indicates that the results are structurally more similar to the ground truth; we implement the evaluation of LPIPS based on AlexNet \cite{krizhevsky2012imagenet}, and smaller LPIPS indicates that the results has a better features similarity between the results and the ground truth.
In addition, we measure the CLIP directional similarity \cite{gal2022stylegan} between image pairs to evaluate the semantic alignment between the editing direction of each method and the target. Specifically, we measure the consistency between the average editing direction from the input images to the generated images and the average direction of the training image pairs. We define the CLIP image directional similarity as follows:
\begin{equation}
1-\operatorname{cos}\left(\Delta_{x \rightarrow y}, \Delta_{x^{\prime} \rightarrow y^{\prime}}\right),
\end{equation}
where $\Delta_{x \rightarrow y}$ is the CLIP direction from the input image to the result image, and $\Delta_{x^{\prime} \rightarrow y^{\prime}}$ is the CLIP direction between the reference images.

\begin{table}[htbp]
\small
\centering
\setlength{\tabcolsep}{0.45mm}{
\begin{tabular}{c|c|c|cc}
\toprule
\multirow{3}{*}{\textbf{Num}} & \multirow{3}{*}{\textbf{Name}} &  \multirow{3}{*}{\textbf{Instruction}} & \multicolumn{2}{c}{\begin{tabular}[c]{@{}c@{}}\textbf{Editing}\\ \textbf{Type}\end{tabular}} \\ \cmidrule{4-5} & & & \textbf{Local} & \textbf{Global} \\ 
\midrule
    1  & boy2girl &"make boy and dog into a girl and cat" & \checkmark  &  
    \\
    2  & midnight &"make it nighttime" &   &\checkmark  
    \\
    3  & sea painting & "turn it into a painting" &   &\checkmark  
    \\
    4  & sketch style & "make the image a pencil sketch" &   &  \checkmark  
    \\
    5  & summer & "make it summer" &   &  \checkmark  
    \\
    6  & wallpaper & "make it snow" &   &  \checkmark  
    \\
    7  & charcoal & "turn it into a charcoal drawing" &   &  \checkmark
    \\
    8  & glasses & "add a pair of glasses" &\checkmark   & 
    \\
    9  & painting & "Make it a painting" &   &  \checkmark 
    \\
    10  & painting snow & "make it snow" &   &  \checkmark  
    \\
    11  & pencil sketch & "as a pencil sketch" &   &  \checkmark
    \\
    12  & purple & "make the sky a deep purple" &   &  \checkmark 
    \\
    13  & snow & "have it snow" &   &  \checkmark  
    \\
    14  & watercolor & "as a watercolor painting" &   &  \checkmark
    \\
    15  & 4dboy & "Turn the boy into a girl" &\checkmark   &    
    \\
    16  & apple & "Turn peaches into apples" & \checkmark  &  
    \\
    17  & cake & "Make it a chocolate cake" & \checkmark   &   
    \\
    18  & cloud kitty & "Make the cat into a bear" &  \checkmark   &   
    \\
    19  & dog2cat & "Make the dog into a cat" & \checkmark   &  
    \\
    20  & juice & "Make it a lemonade" &  \checkmark    &   
    \\
    21  & lava & "Turn it into lava" &   &  \checkmark  
    \\
    22  & rain & "Turn the rain into snow" &   &  \checkmark 
    \\
    23  & read books & "Make newspapers into books" & \checkmark   & 
    \\
    24  & smile & "Add a smile" & \checkmark  &  
    \\
    25  & traffic lights & "make it a heart-shaped light" &\checkmark    &  
    \\
    \bottomrule
\end{tabular}}
\caption{\textbf{Benchmark Presentation.} The benchmark has a total of 25 editing effects, evenly covering both local and global editing.}
\label{tab:benchmark_details}
\end{table}

\section{Benchmark Construction}
\label{supp:benchmark_construction}

\begin{figure}[tb]  \centering\includegraphics[width=1\textwidth]{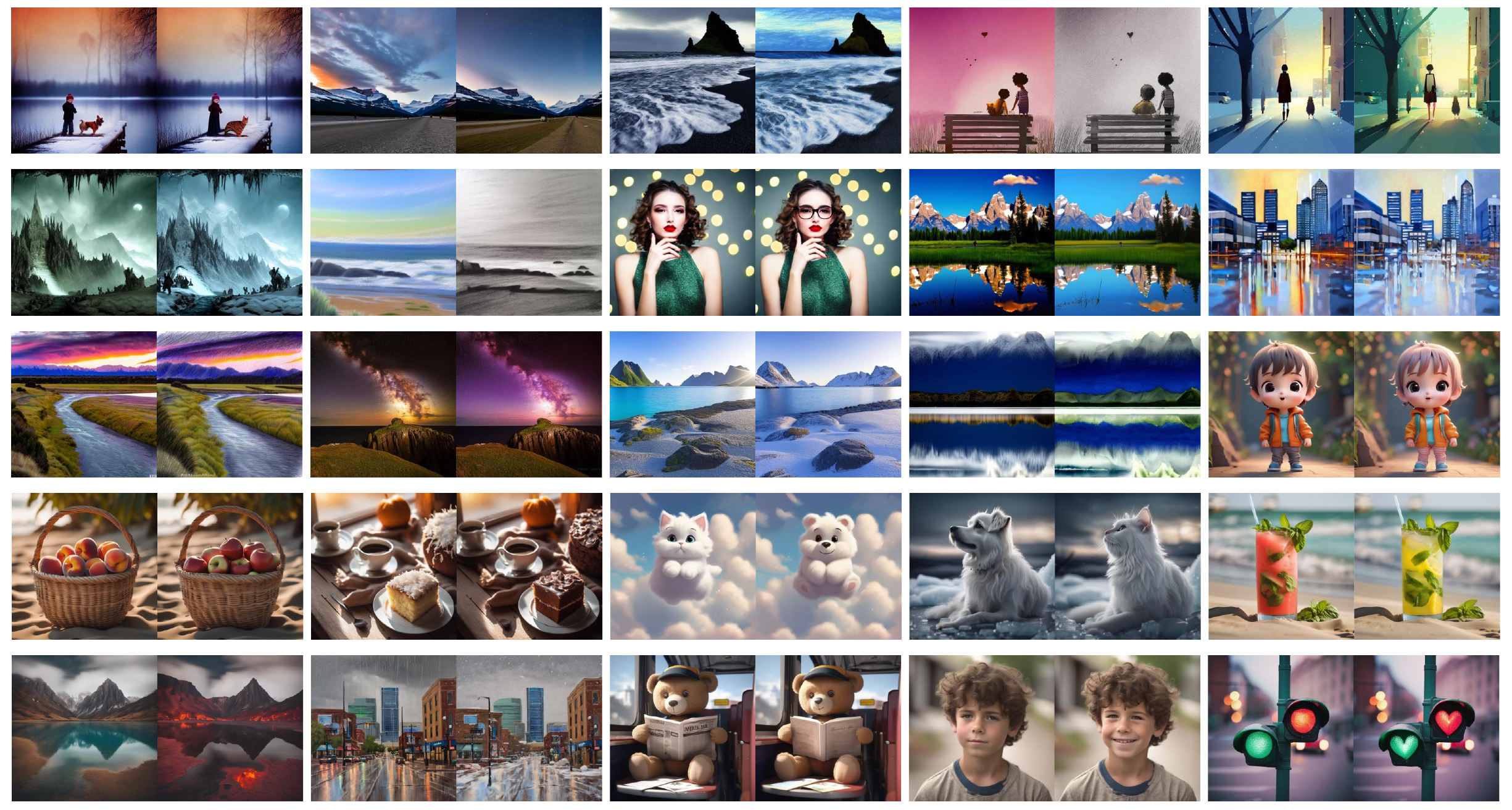}
   \caption{\textbf{Visualization of Our Benchmark.} Our benchmark spans 25 datasets corresponding to different editing effects. It covers a wide range of editing categories and scenarios, allowing for division from multiple dimensions. Each dataset consists of 10 pairs of training images and 5 pairs of testing images, totaling 750 images. We show a pair of before-and-after transformation examples for each editing effect.
   }
   \label{fig:append_benchmark}
\end{figure}

In recent years, there has been rapid development in text-guided image editing methods. The evaluation of image editing effectiveness has also evolved. Initially, the editing effect is solely evaluated through qualitative presentations and user study \cite{hertz2022prompt,mokady2023null}, which led to significant subjectivity. Subsequently, PNP \cite{tumanyan2023plug} establishes a benchmark for text-guided image editing, which assesses the performance of text-based image editing methods using text-image and image-image feature similarity scores. Later, Direct Inversion \cite{ju2023direct} introduces a more robust benchmark for text-guided image editing methods, comprising 700 images and 10 editing types, and utilizes 8 evaluation metrics for an objective and comprehensive assessment. Although these benchmarks are widely used by existing text-guided image editing methods, however, the lack of paired training data prevents them from being applicable to the instruction inversion methods. Visii \cite{nguyen2023visual} utilizes the filtered dataset of IP2P \cite{brooks2023instructpix2pix} for evaluation. However, despite being filtered by CLIP similarity, The overall quality of the IP2P training data is still poor, which is reflected in the quality and fidelity of the images before and after their editing. Furthermore, the dataset of IP2P contains fewer pairs of data for the same editing type, which hinders an accurate assessment of the performance of the instruction inversion method under the few-shot setting. 

To investigate the editing capabilities of various instruction inversion methods in open scenarios and facilitate a fair comparison of these methods, We establish a benchmark named \textit{TOP-Bench} (\textbf{T}ransformation-\textbf{O}riented \textbf{P}aired \textbf{Bench}mark), which can be utilized for both qualitative and quantitative evaluations. 
Our benchmark spans 25 datasets corresponding to different editing effects. It covers a wide range of editing categories and scenarios, allowing for division from multiple dimensions. Each dataset consists of 10 pairs of training images and 5 pairs of testing images, totaling 750 images. Additionally, we provide text instructions aligned with the transformation effects for each dataset.

In order to obtain paired data representing image editing, we refer to the IP2P method of generating data and utilize the existing image editing method P2P \cite{hertz2022prompt} to directly generate paired data before and after editing. For different editing effects, some of them completely replicate the training set of IP2P, i.e., using the image caption as well as the editing instructions are from the training set of IP2P, and the same settings of IP2P are used to generate and filter the high-quality data of the present method so as to represent the editing of the scene in the domain; while for some editing effects, we generate them through the SDXL-based P2P, while the image caption as well as the editing instructions are obtained based on GPT-4 to represent the editing of the out-of-domain scene. 
TOP-Bench provides paired before and after editing data. It is suitable for the evaluation of instruction inversion methods. At the same time, TOP-Bench can be segmented in multiple dimensions to comprehensively evaluate the performance of instruction reversal methods. A detailed presentation of the datasets representing the different editing effects within TOP-Bench and their categorization is shown in Figure \ref{fig:append_benchmark} and Table \ref{tab:benchmark_details}.

\begin{figure}[tb]  \centering\includegraphics[width=1\textwidth]{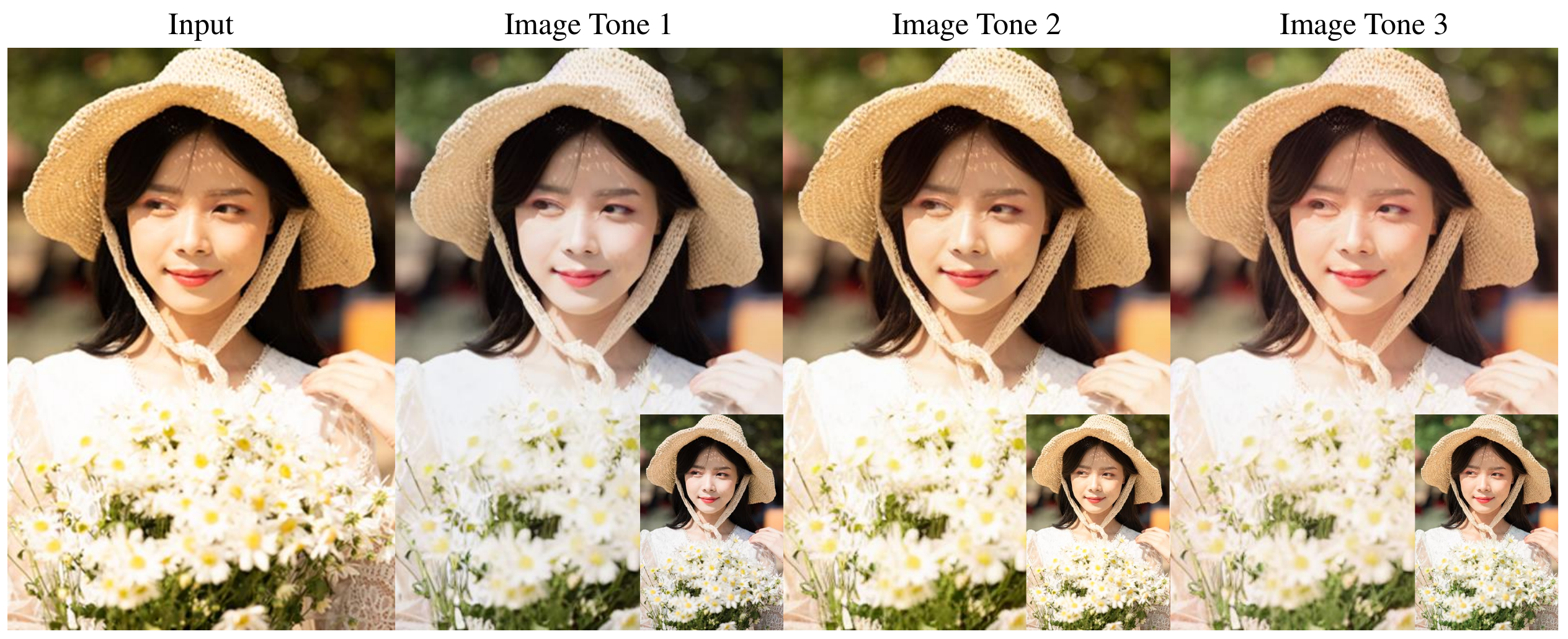}
   \caption{\textbf{Image Tone Modification.} Our InstructBrush can extract different image tones based on a handful of data pairs and apply them to new images. The image on the left gives the input image, and the images on the right show three different outputs, corresponding to three image tones, and the ground truth is given in the bottom right corner of each output image for reference.
   }
   \label{fig:image_tone}
\end{figure}

\section{Extra Applications}
\label{supp:applications}

\noindent \textbf{Image Retouching.}
Image retouching is the process of changing or improving the quality of an image. This involves enhancing colors, removing imperfections, adjusting lighting, or making other edits to improve the overall appearance of an image. Implementing image retouching using the instruction-based image editing models is challenging because the vast majority of image retouching transformations are difficult to describe using textual instructions. Our method helps in this task. Given paired data before and after image retouching, our InstructBrush can extract editing instructions representing this image transformation from the prior of generative model. The aligned instructions obtained through this process facilitate training for downstream tasks. As shown in Figure \ref{fig:image_tone}, our InstructBrush can extract image tones based on few image pairs that represent tonal transformations from PPR10K \cite{liang2021ppr10k} and apply these transformations to new images.

\section{Additional Experiments}
\label{supp:additional_experiments}
\subsection{One-shot Editing}
To further demonstrate the advantages of our method, we test the quantitative results of different methods under 1-shot setting. We use the first pair of images from each training sets within the benchmark as our 1-shot training data pair. All settings were kept the same as in previous experiments. As shown in Table \ref{tab:compare_results_1shot}. Our method outperforms the other methods under 1-shot and has the same trend as the few-shot quantitative experiments. Compared to the results of Visii, our method still shows a more significant improvement on TOP-Local than on TOP-Global for the 1-shot setting. This shows in local editing tasks, training images contain more editing-irrelevant scene information. The content-oriented initialization of Visii introduces them to the initialized instructions, posing a greater obstacle to optimization. Our method can accurately capture the transformations between image pairs and use them for initialization, thus improving instruction generalization.

\begin{table}[t]
\centering
\caption{\textbf{Quantitative Results for One-shot.} We measure the average PSNR, SSIM, LPIPS, and CLIP direction scores of several methods in different editing tasks. In 1-shot settings, our method demonstrates significant superiority over other methods. We highlight in red the percentage of our method that exceeds Visii.}
\resizebox{0.8\columnwidth}{!}{
  \begin{tabular}{clllll}

    \toprule
    Datasets \quad  & Method \quad\quad & PSNR $\uparrow$ & SSIM $\uparrow$  & LPIPS $\downarrow$ & Direct. $\downarrow$ \\ 
    \midrule
    \multirow{3}*{TOP-Global\quad} & IP2P \cite{brooks2023instructpix2pix} \quad\quad  & 16.10 & 0.5467 & 0.2810 & 0.2997 \\ 
                      & Visii \cite{nguyen2023visual} \quad\quad & 16.01 & 0.5071 & 0.3692 & 0.2909 \\ 
                      & \textbf{Ours} \quad\quad & \textbf{17.79}\textcolor{red}{$_{11.1\%\uparrow}$} & \textbf{0.5761}\textcolor{red}{$_{13.6\%\uparrow}$} & \textbf{0.2748}\textcolor{red}{$_{25.6\%\downarrow}$} & 0.3008\\ 
    \midrule 
    \midrule
    \multirow{3}*{TOP-Local\quad} & IP2P \cite{brooks2023instructpix2pix} \quad\quad  & 20.12 & 0.7701 & 0.1648 & 0.5236 \\ 
                      & Visii \cite{nguyen2023visual} \quad\quad & 19.73 & 0.7293 & 0.2309 & 0.5736 \\ 
                      & \textbf{Ours} \quad\quad & \textbf{23.08}\textcolor{red}{$_{17.0\%\uparrow}$} & \textbf{0.8270}\textcolor{red}{$_{13.4\%\uparrow}$} & \textbf{0.1172}\textcolor{red}{$_{49.2\%\downarrow}$} & \textbf{0.4790}\textcolor{red}{$_{16.5\%\downarrow}$}\\ 
    \midrule  
    \midrule
    \multirow{3}*{TOP-Bench\quad} & IP2P \cite{brooks2023instructpix2pix} \quad\quad  & 17.87 & 0.6450 & 0.2298 & 0.3982 \\ 
                      & Visii \cite{nguyen2023visual} \quad\quad & 17.65 & 0.6049 & 0.3083 & 0.4153 \\ 
                      & \textbf{Ours} \quad\quad & \textbf{20.11}\textcolor{red}{$_{13.9\%\uparrow}$} & \textbf{0.6865}\textcolor{red}{$_{13.5\%\uparrow}$} & \textbf{0.2055}\textcolor{red}{$_{33.3\%\downarrow}$} & \textbf{0.3792}\textcolor{red}{$_{8.7\%\downarrow}$}\\ 
    \bottomrule
  \end{tabular}
}
\label{tab:compare_results_1shot}
\end{table}
%

\subsection{More Visualization Results}
\begin{figure}[tb]  \centering\includegraphics[width=1\textwidth]{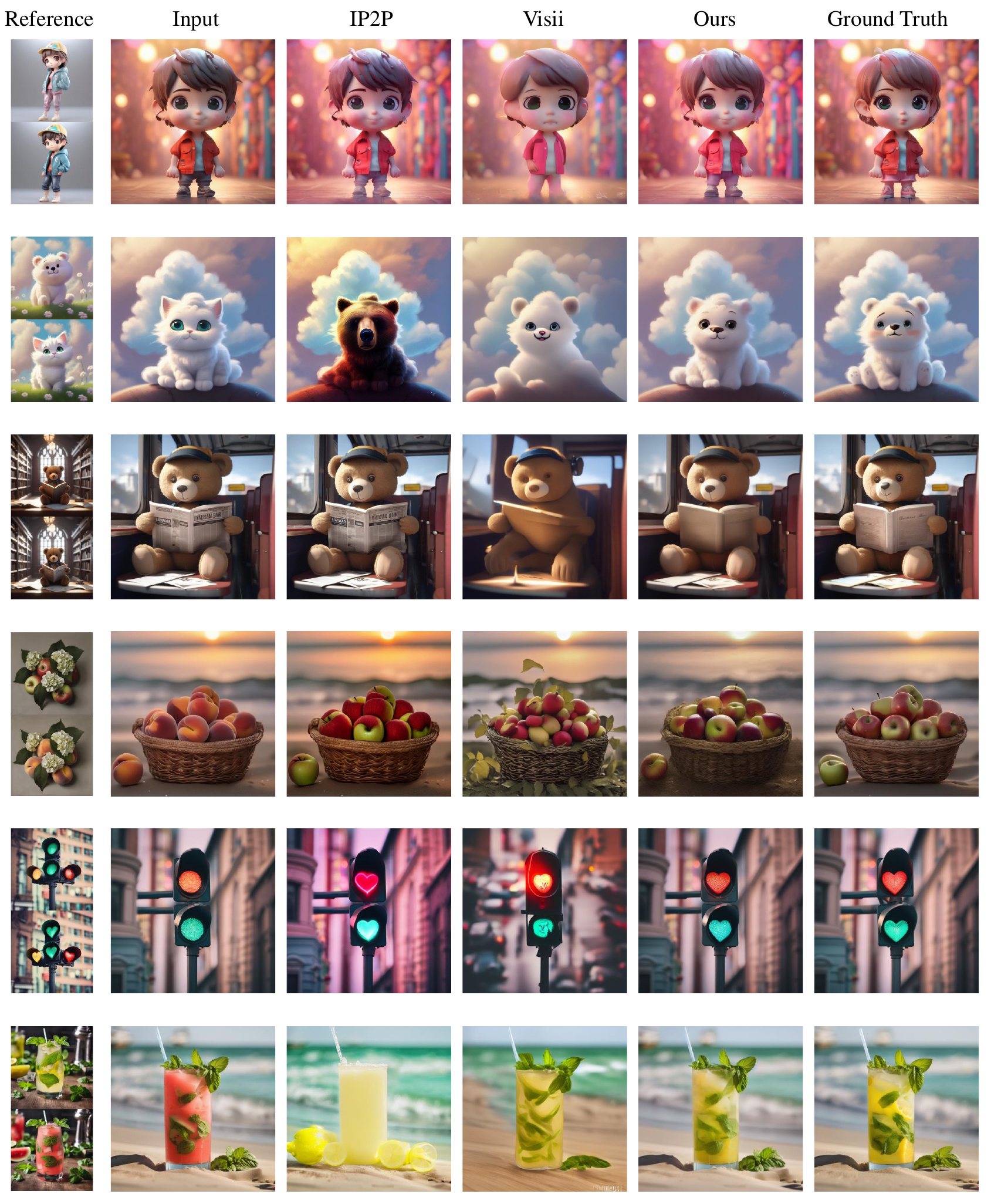}
   \caption{\textbf{Qualitative Comparisons in Local Editing.} We show more qualitative results on local editing. The results show that our method performs well in local editing. It effectively avoids introducing editing-independent information from the training image and shows better instruction generalization.
   }
   \label{fig:append_results1}
\end{figure}

\begin{figure}[tb]  \centering\includegraphics[width=1\textwidth]{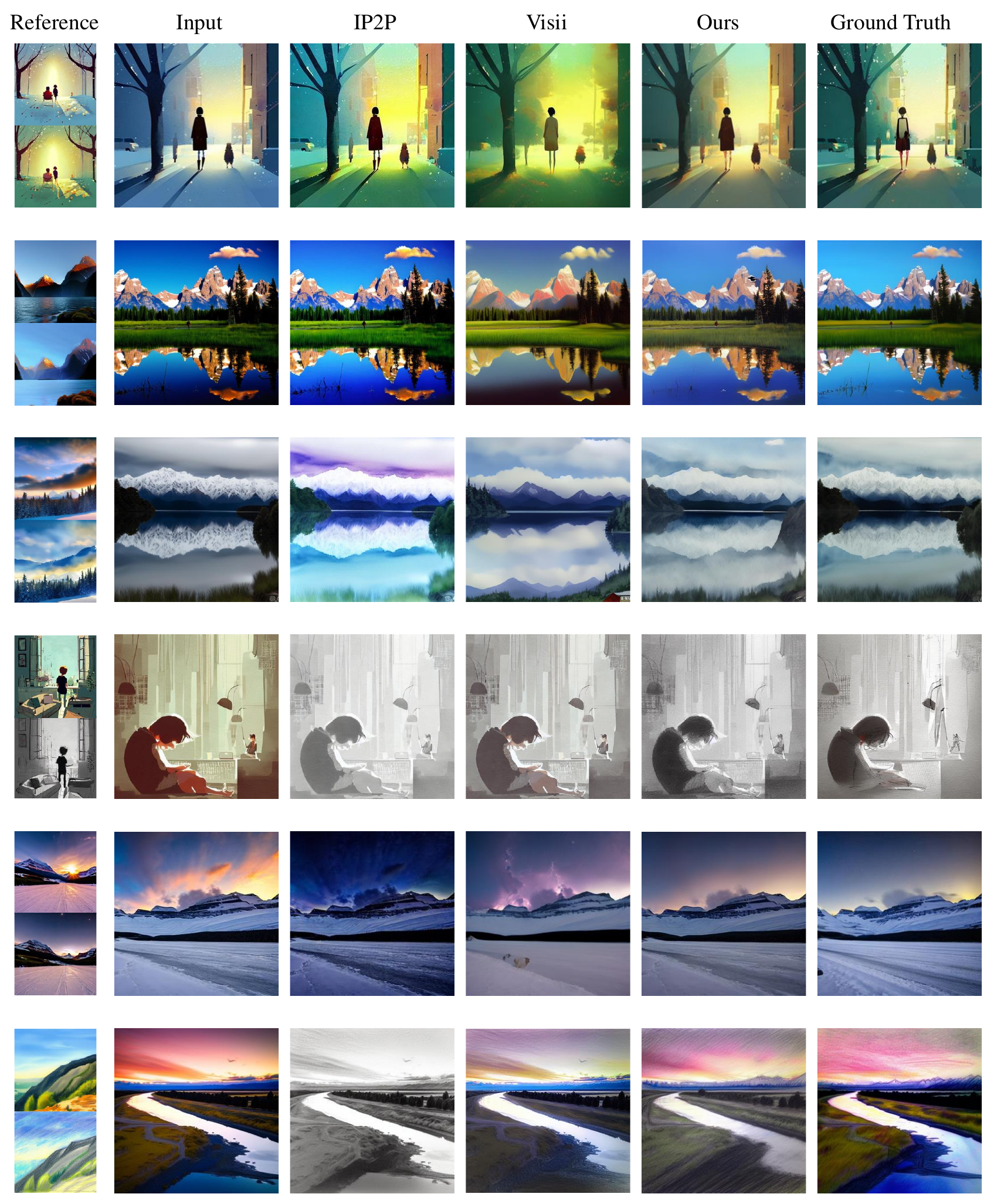}
   \caption{\textbf{Qualitative Comparisons in Global Editing.} We show more qualitative results on global editing. The results show that our method performs well in global editing. It effectively avoids introducing editing-independent information from the training image and shows better instruction generalization.
   }
   \label{fig:append_results2}
\end{figure}

\begin{figure}[tb]  \centering\includegraphics[width=1\textwidth]{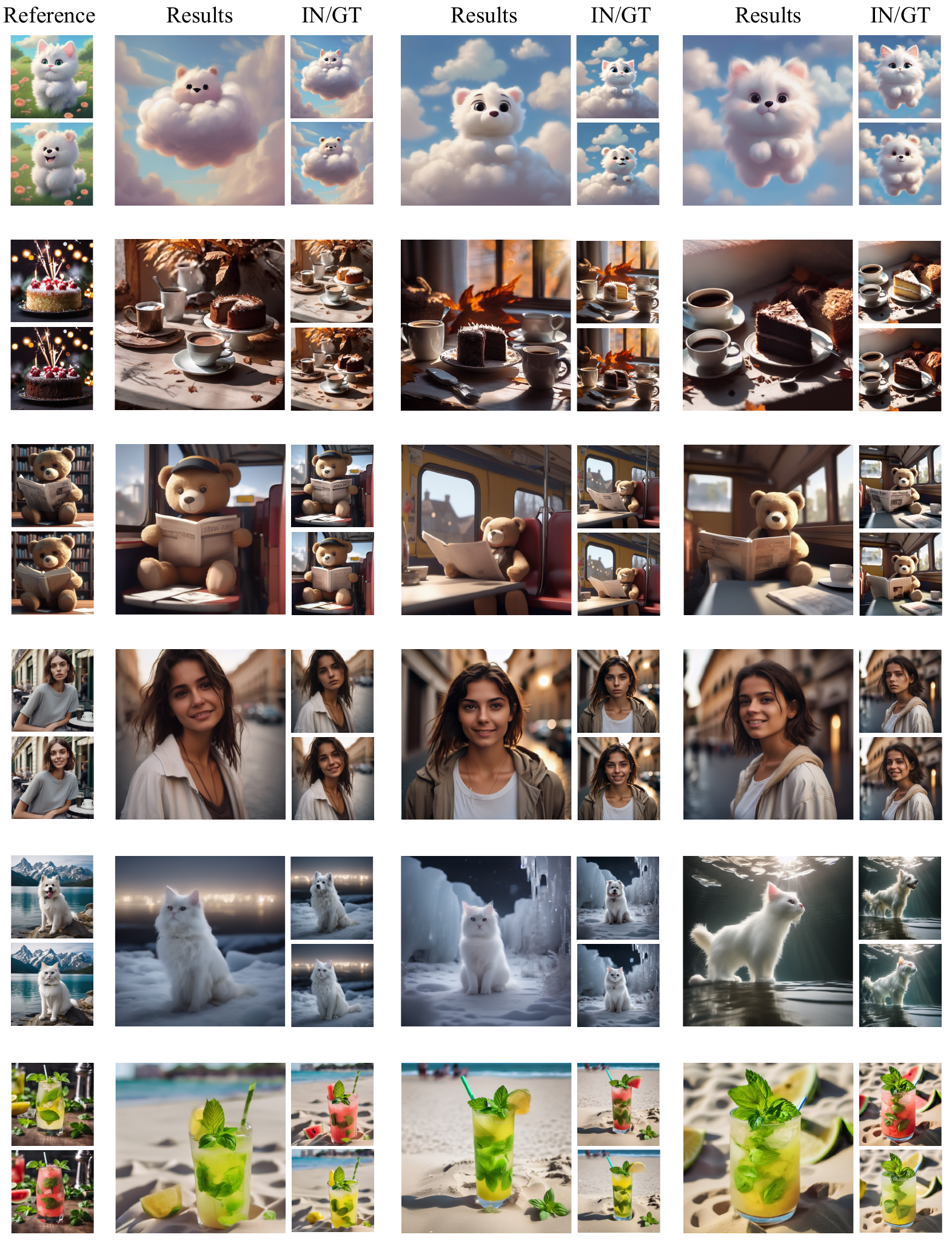}
   \caption{\textbf{More Visualization Results of Our Method for Local Editing.} Our method shows robust performance in local editing. Moreover, it does not introduce the scene information of the training image when editing a new image, which reflects the instructive generality of our method.
   }
   \label{fig:append_results3}
\end{figure}

\begin{figure}[tb]  \centering\includegraphics[width=1\textwidth]{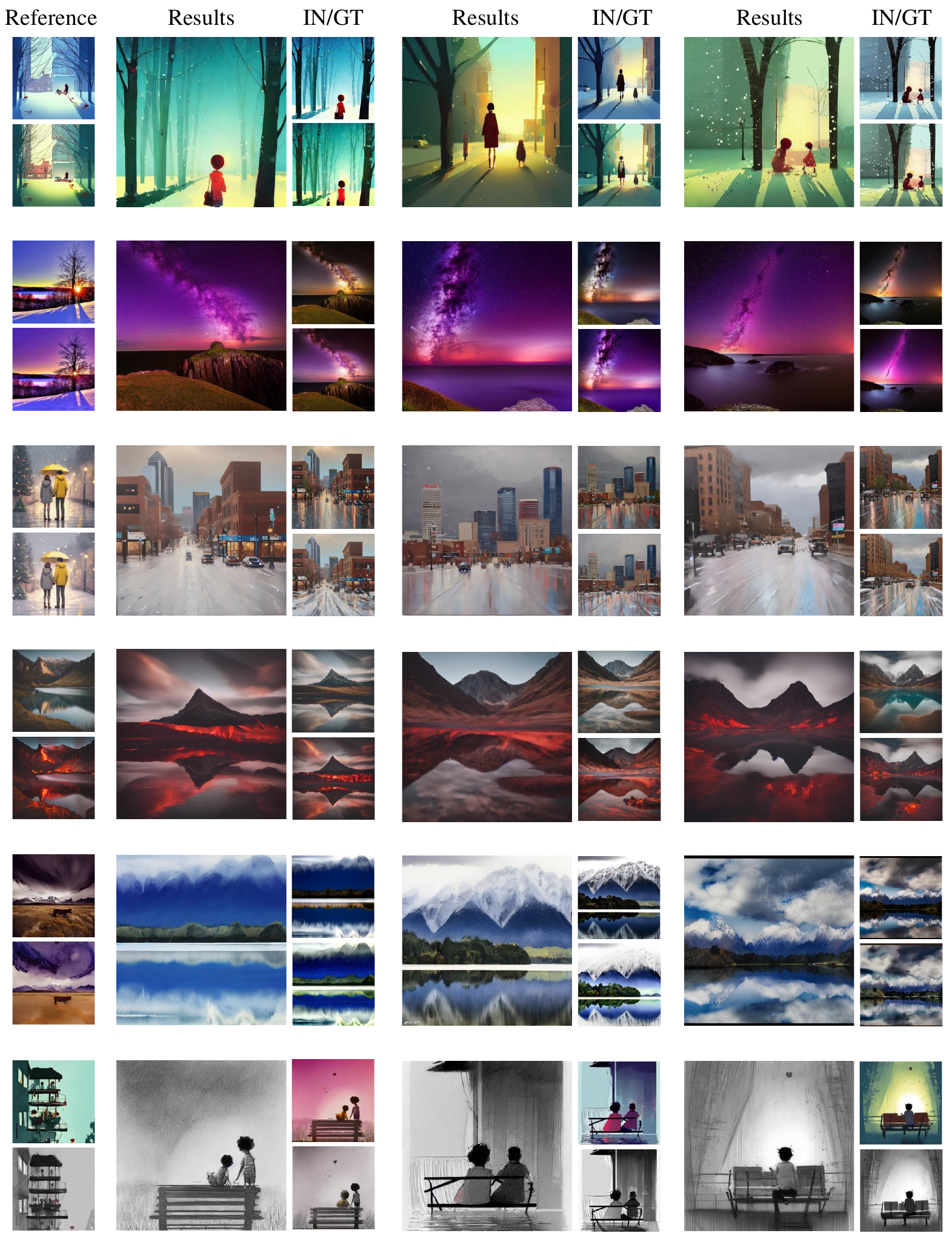}
   \caption{\textbf{More Visualization Results of Our Method for Global Editing.} Our method shows robust performance in global editing. Moreover, it does not introduce the scene information of the training image when editing a new image, which reflects the instructive generality of our method.
   }
   \label{fig:append_results4}
\end{figure}

We show more qualitative comparison results for local and global editing in Figure \ref{fig:append_results1} and Figure \ref{fig:append_results2}, and show more visualization results of our method applied to local and global editing in Figure \ref{fig:append_results3} and Figure \ref{fig:append_results4}. 

